\documentclass[letter]{aa} 

\usepackage{graphicx}
\usepackage{txfonts}

\newcommand{\rasecp}{\mbox{\rlap{.}$^{\rm s}$}} 
\newcommand{\tablenotea}[1]{\parbox{18.3cm}{\indent \footnotesize{#1}}}

\newcommand{\ieee}{IEEE Trans. Antennas Propag.}
\newcommand{\jmst}{J. Mol. Struct.}
\newcommand{\jpca}{J. Phys. Chem. A}
\newcommand{\chemrev}{Chem. Rev.}
\newcommand{\jppa}{J. Photochem. Photobiol. A: Chem.}

\begin{document}

\title{Detection of interstellar HCS and its metastable isomer HSC:\\
new pieces in the puzzle of sulfur chemistry\thanks{Based on observations carried out with the IRAM 30m Telescope. IRAM is supported by INSU/CNRS (France), MPG (Germany) and IGN (Spain).}}

\titlerunning{First detection in space of HCS and HSC}
\authorrunning{Ag\'undez et al.}

\author{
M.~Ag\'undez\inst{1},
N.~Marcelino\inst{1}
J.~Cernicharo\inst{1} \and
M.~Tafalla\inst{2}
}

\institute{
Instituto de F\'isica Fundamental, CSIC, C/ Serrano 123, 28006 Madrid, Spain \and
Observatorio Astron\'omico Nacional (OAN), C/ Alfonso XII 3, 28014 Madrid, Spain
}

\date{Received; accepted}

 
\abstract
{We present the first identification in interstellar space of the thioformyl radical (HCS) and its metastable isomer HSC. These species were detected toward the molecular cloud L483 thanks to observations carried out with the IRAM 30m telescope in the $\lambda$ 3 mm band. We derive beam-averaged column densities of $7\times10^{12}$ cm$^{-2}$ for HCS and $1.8\times10^{11}$ cm$^{-2}$ for HSC, which translate to fractional abundances relative to H$_2$ of $2\times10^{-10}$ and $6\times10^{-12}$, respectively. Although the amount of sulfur locked by these radicals is low, their detection allows to put interesting constraints on the chemistry of sulfur in dark clouds. Interestingly, the H$_2$CS/HCS abundance ratio is found to be quite low, $\sim1$, in contrast with the oxygen analogue case, in which the H$_2$CO/HCO abundance ratio is around 10 in dark clouds. Moreover, the radical HCS is found to be more abundant than its oxygen analogue, HCO. The metastable species HOC, the oxygen analogue of HSC, has not been yet observed in space. These observational constraints are confronted with the outcome of a recent model of the chemistry of sulfur in dark clouds. The model underestimates the fractional abundance of HCS by at least one order of magnitude, overestimates the H$_2$CS/HCS abundance ratio, and does not provide an abundance prediction for the metastable isomer HSC. These observations should prompt a revision of the chemistry of sulfur in interstellar clouds.}


\maketitle

\section{Introduction}

The chemistry of sulfur in interstellar clouds has attracted considerable attention from astronomers in the past decades. For example, which is the main reservoir of sulfur in cold dark clouds continues to be an important open question. A variety of simple sulfur-bearing species is observed in the gas phase of these environments. These comprise the hydride H$_2$S, the carbon-bearing molecules CS, H$_2$CS, C$_2$S, and C$_3$S, the oxides SO and SO$_2$, OCS, the nitrogen-bearing species NS, HNCS, and HSCN, and the ion HCS$^+$ \citep{Adande2010,Agundez2013}. However, these species only account for less than 0.1 \% of the cosmic abundance of sulfur. Moreover, chemical models of cold dark clouds usually need to assume that sulfur is depleted in the gas phase by at least two orders of magnitude to reproduce the abundances derived from observations \citep{Agundez2013}. However, in photodissociation regions such as the Horsehead Nebula, it is found that sulfur is not severely depleted, only by a factor of 3-4 \citep{Goicoechea2006}. The missing sulfur in cold dark clouds is usually thought to be deposited onto dust grains, although it is not clear in which form. Some sulfur could be trapped in the core of grains as refractory compounds such as FeS or MgS, although it is not known how much sulfur could be in this form. Unfortunately, the depletion factor of sulfur in diffuse interstellar clouds is uncertain because the S$^+$ lines are often saturated \citep{Jenkins2009}. Ices could also be an important reservoir of sulfur in cold dark clouds. Solid OCS and SO$_2$ are thought to be responsible of absorption features in the infrared spectra of young stellar objects, with inferred abundances of 0.2-0.5 \% and 0.6-6 \%, respectively, of the available sulfur \citep{Palumbo1997,Boogert1997}. It has been proposed that other ices such as polysulfanes could be a major reservoir of sulfur \citep{Jimenez-Escobar2011,Druard2012}. It is also possible that gas-phase molecules not yet identified could account for a significant fraction of this element.

The missing sulfur problem has attracted renewed attention in recent years. New potentially important sulfur reservoirs have been explored. For example, the S$_2$-containing molecules H$_2$S$_2$, HS$_2$, and S$_2$ have been unsuccessfully searched for toward the low-mass protostar IRAS\,16293-2422 by \cite{Martin-Domenech2016}, while the detection of gas-phase H$_2$S in the shocked region L1157-B1 by \cite{Holdship2016} led these authors to conclude that a significant fraction of sulfur must be in the form of H$_2$S ice. The chemistry of sulfur in dark clouds has been recently revisited observationally by \cite{Fuente2016} and theoretically by \cite{Vidal2017}. These latter authors point to gas-phase atomic sulfur or SH and H$_2$S ices as main sulfur reservoirs. It is also worth noting that two new sulfur-bearing molecules have been recently detected in space: HS$_2$ has been identified in the Horsehead Nebula \citep{Fuente2017} and the ion NS$^+$ has been ubiquitously found in dense molecular clouds \citep{Cernicharo2018}.

In this Letter we present the first detection in interstellar space of two new sulfur-bearing molecules: the radical HCS and its metastable isomer HSC. These species have been identified in the dense cloud L483, a source that hosts a rich chemistry and where we have recently discovered new molecules like the ketenyl radical (HCCO), protonated cyanogen (NCCNH$^+$), and the ion NS$^+$ \citep{Agundez2015a,Agundez2015b,Cernicharo2018}.

\begin{table*}
\caption{Observed line parameters of HCS and HSC in L483.} \label{table:lines}
\centering
\begin{tabular}{llccccccc}
\hline \hline
\multicolumn{1}{c}{Species} & \multicolumn{1}{c}{Transition} & \multicolumn{1}{c}{Frequency} & \multicolumn{1}{c}{$E_u$} & \multicolumn{1}{c}{$A_{ul}$} & \multicolumn{1}{c}{$g_u$} & \multicolumn{1}{c}{$V_{\rm LSR}$} & \multicolumn{1}{c}{$\Delta v$}      & \multicolumn{1}{c}{$\int T_A^* dv$} \\
                                              &  & \multicolumn{1}{c}{(MHz)}                 & \multicolumn{1}{c}{(K)}                 & \multicolumn{1}{c}{(s$^{-1}$)}                 & & \multicolumn{1}{c}{(km s$^{-1}$)}    & \multicolumn{1}{c}{(km s$^{-1}$)} & \multicolumn{1}{c}{(mK km s$^{-1}$)} \\
\hline
HCS & $2_{0,2}-1_{0,1}$~~$J=5/2-3/2$~~$F=3-2$  & 80553.516 & 5.8 & $4.50\times10^{-7}$ & 7 & +5.43(3)   & 0.61(5)   & 10.5(9)  \\ 
        & $2_{0,2}-1_{0,1}$~~$J=5/2-3/2$~~$F=2-1$  & 80565.596 & 5.8 & $4.34\times10^{-7}$ & 5 & +5.33(5)   & 0.77(12) & 10.4(12) \\ 
        & $2_{0,2}-1_{0,1}$~~$J=3/2-1/2$~~$F=2-1$  & 80596.409 & 5.8 & $3.41\times10^{-7}$ & 5 & +5.56(6)   & 0.88(12) &   9.0(11) \\ 
        & $2_{0,2}-1_{0,1}$~~$J=3/2-1/2$~~$F=1-0$  & 80611.994 & 5.8 & $2.51\times10^{-7}$ & 3 & +5.32(10) & 0.70(20) &   3.5(11) \\ 
        & $2_{0,2}-1_{0,1}$~~$J=3/2-1/2$~~$F=1-1$  & 80618.820 & 5.8 & $1.79\times10^{-7}$ & 3 & +5.67(7)   & 0.36(23) &   2.4(10) \\ 
\hline
HSC & $2_{0,2}-1_{0,1}$~~$J=5/2-3/2$~~$F=2-1$  & 81192.825 & 5.8 & $1.55\times10^{-5}$ & 5 & --              &   --          &    -- $^a$ \\ 
        & $2_{0,2}-1_{0,1}$~~$J=3/2-1/2$~~$F=2-1$  & 81194.075 & 5.9 & $1.16\times10^{-5}$ & 5 & +5.39(8)   & 0.65(15) &   4.5(11)  \\ 
        & $2_{0,2}-1_{0,1}$~~$J=5/2-3/2$~~$F=3-2$  & 81199.988 & 5.9 & $1.55\times10^{-5}$ & 7 & +5.42(5)$^b$ & 0.43(9)$^b$   &   4.0(8)$^b$   \\ 
\hline
\end{tabular}
\tablenotea{\\
Numbers in parentheses are 1$\sigma$ uncertainties in units of the last digits. The antenna temperature scale can be converted to main beam brightness temperature by dividing by ($B_{\rm eff}/F_{\rm eff}$), which is 0.82/0.95 for the IRAM 30m telescope at 81 GHz. $^a$ Line is not visible because it overlaps with a frequency switching negative artifact of the HSC 2$_{0,2}$-1$_{0,1}$ $J=5/2-3/2$ $F=3-2$ line at 81199.988 MHz (see Fig.~\ref{fig:hsc_lines}). $^b$ Intensity is reduced because line overlaps with a frequency switching negative artifact of the HSC 2$_{0,2}$-1$_{0,1}$ $J=5/2-3/2$ $F=2-1$ line at 81192.825 MHz, line width is likely reduced as well because the line partially overlaps with a negative artifact of a line of CH$_2$CN lying 7.36 MHz higher in frequency (see Fig.~\ref{fig:hsc_lines}), and the derived $V_{\rm LSR}$ is probably affected as well.
}
\end{table*}

\section{Observations}

The observations were carried out with the IRAM 30m telescope in the frame of a $\lambda$ 3 mm line survey of L483. We adopted the coordinates of the dense core, $\alpha_{2000.0}$ = 18$^{\rm h}$17$^{\rm m}$29\rasecp8, $\delta_{2000.0}$ = $-04^{\circ}$39$'$38$''$, which correspond to the position of the infrared source IRAS\,18148$-$0440 \citep{Fuller1993}. This position coincides also with the maximum of intensity of CH$_3$OH emission \citep{Tafalla2000}. We used the EMIR receiver E090 in single sideband mode, with image rejections $>$ 10 dB, and employed the frequency-switching technique with a frequency throw of 7.2 MHz. We used the FTS backend in its narrow mode, providing a bandwidth of 1.8 GHz and a spectral resolution of 50 kHz, which translates to velocity resolutions of 0.19-0.13 km\,s$^{-1}$ in the 80-116 GHz frequency range. The intensity scale is calibrated using two absorbers at different temperatures and the atmospheric transmission model ATM \citep{Cernicharo1985,Pardo2001}. We express intensities in terms of $T_A^*$, the antenna temperature corrected for atmospheric absorption and for antenna ohmic and spillover losses. The uncertainty in $T_A^*$ is estimated to be around 10 \%.

The observations around 81 GHz, where the lines of HCS and HSC presented here lie, were carried out in two observing sessions, during November 2016 and December 2017. Weather conditions were fairly good during these two observing runs, with amounts of precipitable water vapor in the ranges 2.1-5.8 mm and 1.5-2.8 mm, respectively, leading to system temperatures of 85-112 K and 75-92 K, respectively, for elevation angles above 30$^{\circ}$. The telescope focus was checked on planets at the beginning of each observing session, which typically lasted $\sim5$~h. The telescope pointing was regularly checked every one hour and half by observing the nearby radio source 1741$-$038. Pointing errors were typically 2-3$''$. The beam size of the IRAM 30m telescope at 81 GHz is 30$''$. The on source integration times were 5.05 h in November 2016 and 4.12 h in December 2017. After including data of the two observing sessions and averaging spectra of horizontal and vertical polarizations, the resulting spectrum is very sensitive, with a $T_A^*$ rms noise level of $\sim$1.5 mK per 50 kHz channel in the spectral region around 81 GHz. The lines of H$_2$CS at 101.5 GHz, 103.0 GHz, and 104.6 GHz were observed in May 2017 in the frame of the same $\lambda$ 3 mm line survey of L483. Weather conditions were also good, with similar numbers to those given above. More details on these observations will given elsewhere.

\begin{figure}
\centering
\includegraphics[angle=0,width=\columnwidth]{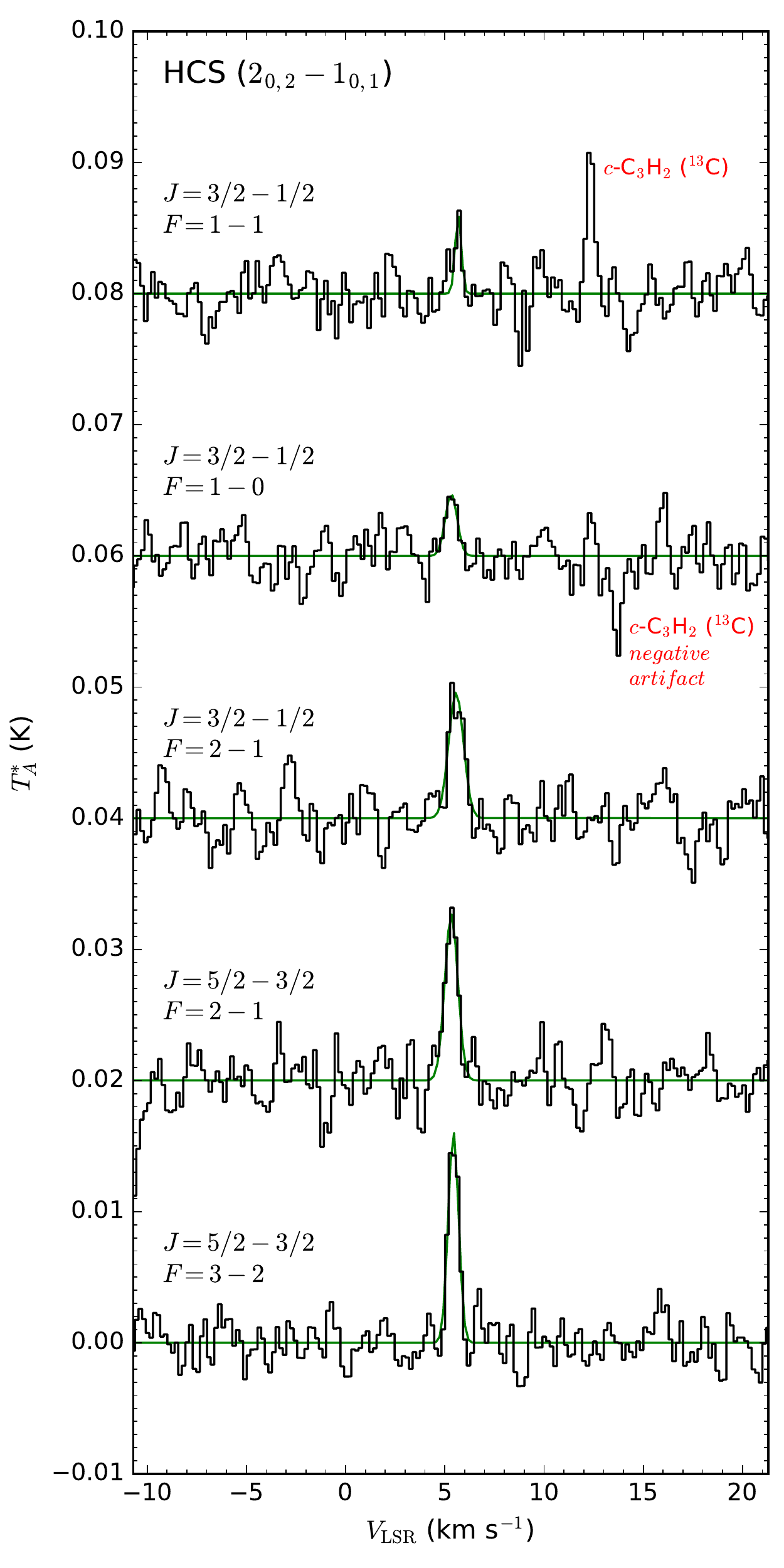}
\caption{Lines of HCS observed in L483 are shown from bottom to top in order of increasing frequency. Note that the observed line intensity decreases as the frequency increases.} \label{fig:hcs_lines}
\end{figure}

\begin{figure}
\centering
\includegraphics[angle=0,width=\columnwidth]{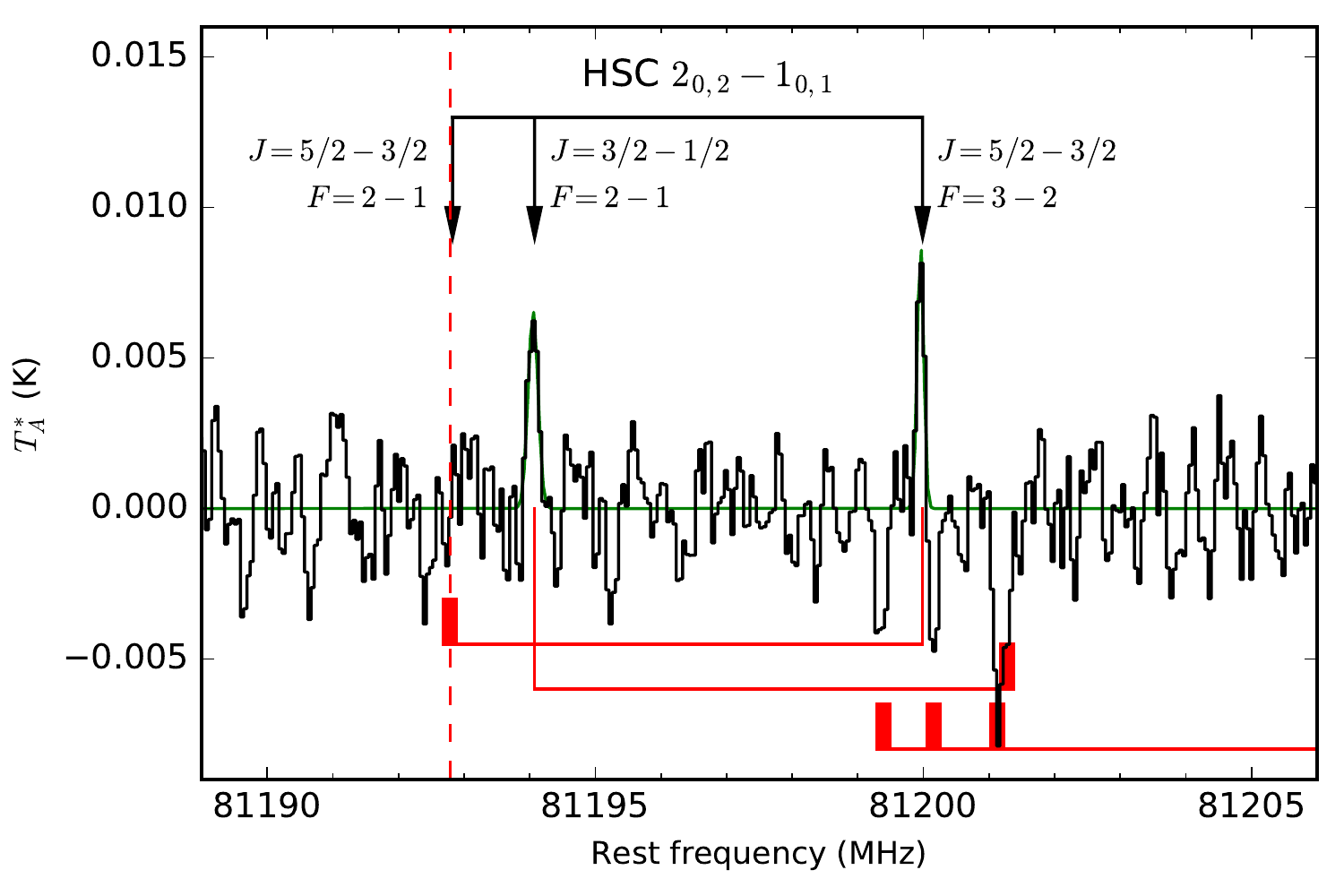}
\caption{Spectra of L483 around 81.2 GHz showing the lines assigned to HSC. The rest frequency scale corresponds to a LSR systemic velocity of +5.3 km\,s$^{-1}$. The bottom vertical red marks indicate the position of frequency switching negative artifacts. Note that the $J=5/2-3/2$ $F=2-1$ line is not visible because it is artificially cancelled with a frequency switching negative artifact, while the $J=5/2-3/2$ $F=3-2$ is also affected by negative artifacts (see text).} \label{fig:hsc_lines}
\end{figure}

\section{Results}

The spectrum of L483 around 80.5 GHz shows a series of emission lines which can be unambiguously assigned to the strongest components of the $2_{0,2}-1_{0,1}$ rotational transition of the radical HCS. In the spectral region around 81.2 GHz there are two emission lines whose frequencies precisely coincide with two of the strongest components of the $2_{0,2}-1_{0,1}$ rotational transition of the metastable isomer HSC (see Table~\ref{table:lines} and Figs.~\ref{fig:hcs_lines} and \ref{fig:hsc_lines}). These species have not been previously observed in space.

The radicals HCS and HSC have a bent structure with a $^2A'$ ground electronic state. The HSC isomer is $\sim40$ kcal mol$^{-1}$ less stable than HCS \citep{Puzzarini2005}. Their rotational levels $N_{K_a,K_c}$ split in a fine (electronic spin-rotation interaction) and hyperfine (H nuclear spin) structure described by the quantum numbers $J$ and $F$, respectively. The rotational spectra of the two radicals have been characterized in the laboratory \citep{Habara2002,Habara2000}, although in the case of the more stable isomer HCS only transitions with $K_a=0$ were observed. Line frequencies were obtained from the Cologne Database for Molecular Spectroscopy\footnote{See \texttt{http://www.astro.uni-koeln.de/cdms/}} \citep{Muller2005}. The total dipole moments of HCS and HSC have been calculated as 0.96 D and 2.63 D, respectively, with components along the $a$ axis of 0.430 D and 2.493 D, respectively \citep{Puzzarini2005}. We use these latter values for the computation of the column densities because the observed lines correspond to $a$-type transitions.

The identification of HCS in L483 relies on the detection of a quintet of lines corresponding to the five strongest fine and hyperfine components of the $2_{0,2}-1_{0,1}$ rotational transition. The line parameters are given in Table~\ref{table:lines} and the lines are shown in Fig.~\ref{fig:hcs_lines}. These five transitions have line strengths that decrease with increasing frequency (see Einstein coefficients in Table~\ref{table:lines}), which is consistent with the observed pattern of line intensities. The brightest line, lying at 80553.516 MHz, has a peak intensity of $T_A^*\sim15$ mK and is well detected above the noise level, with a signal to noise ratio (SNR) $\gtrsim$ 10$\sigma$, while the faintest one, lying at 80618.820 MHz, is observed with a peak intensity of $T_A^*\sim5$ mK and is detected at a lower confidence level, $\sim3\sigma$ (see Fig.~\ref{fig:hcs_lines}). In the case of the metastable isomer HSC, the $2_{0,2}-1_{0,1}$ rotational transition has also multiple components arising from the fine and hyperfine structure. The three strongest ones are those listed in Table~\ref{table:lines}. Two of these are clearly detected in the spectrum of L483 at confidence levels of 4-5$\sigma$, although the one lying at 81192.825 MHz is not visible (see Fig.~\ref{fig:hsc_lines}). The most likely reason is that the frequency separation between the components at 81192.825 MHz and 81199.988 MHz is similar to the frequency throw of 7.2 MHz adopted in the frequency switching observations. Therefore, it accidentally happens that each line overlaps with a frequency switching negative artifact of the other line. As a result, in our observed spectrum, the weaker line at 81192.825 MHz is artificially cancelled and the intensity of the stronger line at 81199.988 MHz is also reduced.

Assuming a rotational temperature of 10 K, close to the gas kinetic temperature \citep{Fuller1993,Anglada1997}, we derive beam-averaged column densities of $7\times10^{12}$ cm$^{-2}$ and $1.8\times10^{11}$ cm$^{-2}$ for HCS and HSC, respectively, in L483. These values translate to fractional abundances relative to H$_2$ of $2\times10^{-10}$ for HCS and $6\times10^{-12}$ for HSC, adopting the value of $N$(H$_2$) of $3 \times 10^{22}$ cm$^{-2}$ derived by \cite{Tafalla2000} from observations of C$^{17}$O. Note that the column density of HCS only includes $K_a=0$ levels. Using the rotational constants A, B, and C from \cite{Habara2002}, we estimate that the contribution of $K_a\neq0$ levels to the partition function at 10 K is just 2\%. Therefore, the HCS/HSC column density ratio in L483 is $\sim$40. The related molecule H$_2$CS was also observed in the course of the $\lambda$ 3 mm line survey of L483. For H$_2$CS we estimate excitation temperatures in the range 5-10 K, adopting a volume density of H$_2$ of $3.4\times10^4$ cm$^{-3}$ \citep{Jorgensen2002} and taking into account that H$_2$CS should be excited through collisions more efficiently than H$_2$CO due to the larger geometrical cross section. From the observed intensities and assuming rotational temperatures in the range 5-10 K, we derive a beam-averaged column density of $(1.0-2.1)\times10^{13}$ cm$^{-2}$ for H$_2$CS.

Therefore, the radical HCS has an abundance similar to that of thioformaldehyde, with a H$_2$CS/HCS column density ratio of 1.4-3, i.e., of the order of unity. The situation here is markedly different from that of the oxygen-bearing analogues. In L483, the H$_2$CO/HCO column density ratio is 19 \citep{Tafalla2000,Agundez2015a}, in line with the values found in other cold dark clouds, which are in the range 7-22 (see \citealt{Ocana2017} and references therein). Therefore, while the radical HCO is around ten times less abundant than H$_2$CO in cold dark clouds, in L483 the HCS radical is as abundant as H$_2$CS. Moreover, HCS is even more abundant than HCO in L483. The presence of the metastable isomer HSC in cold dark clouds makes also a difference between oxygen and sulfur chemistries. The oxygen analogue HOC has not been observed in space, essentially because there is a complete lack of experimental information for this species. Both metastable isomers HOC and HSC lie $\sim$40 kcal mol$^{-1}$ above their respective stable isomers, HCO and HCS. However, HOC lies also above the fragments H and CO, whereas HSC is more stable than H and CS \citep{Marenich2003,Puzzarini2005,Perez-Juste2007}. Still, the metastable isomer HOC should be stable enough to be characterized because the decay to H and CO is predicted to have a barrier \citep{Marenich2003}. The search for HOC in space must in any case await its detection in the laboratory.

\section{Discussion}

The fractional abundances relative to H$_2$ derived for HCS and HSC, a few 10$^{-10}$ and several 10$^{-12}$, respectively, indicate that these species do not lock an important fraction of sulfur. However, the finding that HCS has an abundance similar to H$_2$CS and the detection of the metastable isomer HSC put interesting constraints on the chemistry of sulfur in cold dark clouds.

The chemistry of sulfur in dark clouds has been recently revisited by \cite{Vidal2017} in a study in which the sulfur chemical network was thoroughly revised. Interestingly, this study discusses the potential detectability of HCS based on the non-negligible abundance predicted for this radical. The calculated abundance between 10$^5$ and 10$^6$ yr is however in the range 10$^{-12}$-10$^{-11}$ relative to H$_2$, which is below the observed value by at least one order of magnitude. Moreover, the observational finding of a H$_2$CS/HCS abundance ratio close to 1 is not well accounted for by their model, which predicts that H$_2$CS is one to two orders of magnitude more abundant than HCS. No prediction is made for the metastable isomer HSC as this species is not included in their model.

The main reactions leading to HCS in the model of \cite{Vidal2017} are
\begin{equation}
\rm S + CH_2 \rightarrow HCS + H, \label{reac:s+ch2}
\end{equation}
\begin{equation}
\rm H_2CS^+ + e^- \rightarrow HCS + H, \label{reac:h2cs+}
\end{equation}
\begin{equation}
\rm H_3CS^+ + e^- \rightarrow HCS + H + H, \label{reac:h3cs+}
\end{equation}
\begin{equation}
\rm C + H_2S \rightarrow HCS + H. \label{reac:c+h2s}
\end{equation}
Reactions~(\ref{reac:s+ch2}), (\ref{reac:h2cs+}), and (\ref{reac:h3cs+}) are the sulfur analogues of the main reactions forming the HCO radical in cold dark clouds, in which case chemical models calculate an abundance in good agreement with observed values \citep{Agundez2015a,Bacmann2016,Ocana2017}. Observations however indicate that the oxygen and sulfur cases behave differently, as indicated by the different abundance ratios: H$_2$CO/HCO $>1$ and H$_2$CS/HCS $\sim1$. Reaction~(\ref{reac:s+ch2}) is basically considered by similarity with the O + CH$_2$ reaction, which has been suggested to yield HCO. Note however that it is not fully clear whether this reaction can proceed \citep{Ocana2017}, and thus it would be of great interest to study theoretically the reactions of O and S atoms with CH$_2$ to evaluate their role in the formation of HCO and HCS, respectively, and of their metastable isomers HOC and HSC. Reaction~(\ref{reac:h2cs+}) is a feasible route to HCS, although since the most stable structure of the ion H$_2$CS$^+$ has the two hydrogen atoms bonded to carbon \citep{Curtiss1992}, it would require an important rearrangement to yield HSC. Reaction~(\ref{reac:h3cs+}) is a likely pathway to HCS. Moreover, the similar abundances derived for H$_2$CS and HCS are suggestive of a common origin and thus the dissociative recombination of H$_3$CS$^+$ could be a common source of both species. Reaction~(\ref{reac:c+h2s}) has been studied experimentally and theoretically  and offers an interesting route to both HCS and HSC. The reaction is rapid at room temperature, with a measured rate constant of $2.5\times10^{-10}$ cm$^{3}$ s$^{-1}$ \citep{Deeyamulla2006}. Theoretical calculations indicate that both HCS and HSC can form without energy barrier \citep{Ochsenfeld1999,Galland2001}, although the formation of HCS is more exothermic (40-44 kcal mol$^{-1}$) that that of HSC (just 2-4 kcal mol$^{-1}$), which could favor HCS over HSC. In fact, this would be in agreement with crossed-beam experiments \citep{Kaiser1999,Ochsenfeld1999}. Note that although reactions~(\ref{reac:s+ch2}-\ref{reac:c+h2s}) are plausible routes to HCS, their implementation in the model of \cite{Vidal2017} is not enough to rise the abundance of this radical up to the level observed in L483. It is clear that the chemistry of sulfur still needs to be revised in the light of the new observational constraints provided by the detection of HCS and HSC.

\section{Conclusions}

We have presented the first detection in space of the thioformyl radical (HCS) and its metastable isomer HSC. These species have been observed in the molecular cloud L483, with fractional abundances relative to H$_2$ of a few 10$^{-10}$ for HCS and various 10$^{-12}$ for HSC. The fraction of sulfur locked by these radicals is low. However, their detection puts interesting constraints on the chemistry of sulfur in dark clouds. We find significant differences with respect to the oxygen analogue species. First, the observed H$_2$CS/HCS abundance ratio is close to unity, while the H$_2$CO/HCO abundance ratio in dark clouds is found to be around 10. Second, the metastable isomer HSC is found with a low abundance, while its oxygen analogue HOC has not been yet observed in space. The latest chemical models of dark clouds cannot account for the relative abundance of HCS, the H$_2$CS/HCS ratio, and the presence of HSC, suggesting that a revision of the chemistry of sulfur is required.

\begin{acknowledgements}

We thank the referee for a careful reading of the manuscript and the IRAM 30m staff for their help during the observations. We acknowledge funding support from the European Research Council (ERC Grant 610256: NANOCOSMOS) and from Spanish MINECO through grant AYA2016-75066-C2-1-P. M.A. also thanks funding support from the Ram\'on y Cajal programme of Spanish MINECO (RyC-2014-16277).

\end{acknowledgements}

\end{document}